\documentclass[%
 reprint,
 amsmath,amssymb,
 aps,
]{revtex4-1}

\usepackage[colorlinks = true,
            linkcolor = blue,
            urlcolor  = blue,
            citecolor = blue,
            anchorcolor = blue]{hyperref}
\usepackage{graphicx}% Include figure files
\usepackage{dcolumn}% Align table columns on decimal point
\usepackage{bm}% bold math
\usepackage[toc,page]{appendix}
\DeclareMathOperator{\erfi}{erfi}
\DeclareMathAlphabet{\pazocal}{OMS}{zplm}{m}{n}
\newcommand \Pu{\mathcal{P}}
\begin{document}

%\preprint{APS/123-QED}

\title{Magneto-Vortical evolution of QGP in heavy ion collisions}%: 2+1 dimensional analytic solution }% Force line breaks with \\

\author{Ashutosh Dash}
\author{Victor Roy}
\author{Bedangadas Mohanty}
 \email{ashutosh.dash@niser.ac.in}
 \email{victor@niser.ac.in}
 \email{bedanga@niser.ac.in}
\affiliation{%
 School of Physical Sciences, National Institute of Science Education
and Research, HBNI, Jatni - 752050, Odisha, India.\\
}%

\date{\today}% It is always \today, today,

\begin{abstract}
The interplay of magnetic field and thermal vorticity in a relativistic ideal fluid might 
generate fluid vorticity during the fluid evolution 
%give rise to breaking of topological constraints (or generation of fluid vorticity) 
provided the flow fields and the entropy density of the fluid is inhomogeneous \cite{Mahajan:2010}.
Exploiting this fact and assuming large magnetic Reynolds number we study the evolution of 
generalised magnetic field ($\hat{B}$) which is defined as a combination of 
the usual magnetic field ($\vec{B}$) and relativistic thermal vorticity ($\omega^{\mu\nu}$),
in a 2(space)+1(time) dimensional isentropic evolution of Quark Gluon Plasma
(QGP) with longitudinal boost invariance. The temporal evolution of  $\hat{B}$
is found to be different than $\vec{B}$ , and the $\hat{B}$ evolution also
depends on the position of the fluid along the beam direction
(taken along the z axis) with respect to the mid-plane $z=0$.
Further it is observed that the transverse components
($\hat{B_{x}}$, $\hat{B_{y}}$) evolve differently around the mid-plane. 
\iffalse
We study the time evolution of generalaised magnetic field by combining the usual magnetic field with the fluid vorticity
in a relativistic framework.The demands of special relativity break the constraints imposed by the frozen-in theorem and Kelvin's circulation
theorem by the inclusion of a non-trivial source term that depends on the gradients of entropy and flow fields, given they
are inhomogeneous.
\fi
\begin{description}
\item[PACS numbers]
  25.75.-q, 12.38.Mh, 25.75.Ag
\end{description}
\end{abstract}

\pacs{Valid PACS appear here}% PACS, the Physics and Astronomy
\maketitle

%\tableofcontents

%%%%%%%%%%%%%%%%%%%%%%%%%%%%%%%%%%%%%%%%%%%%%%%%%%%%%%%%%%%%%%%%%%%%
%% Introduction

\section{\label{sec:intro}Introduction}

Understanding the space-time evolution of intense electromagnetic
field produced in the initial stage of heavy ion collisions is an
important subject, particularly the interplay of magnetic field and
quantum anomalies, which includes  the chiral magnetic effect~\cite{Kharzeev:2007jp,Fukushima:2008xe}, 
chiral magnetic wave~\cite{Burnier:2011bf}, chiral vortical
effect~\cite{Kharzeev:2010gr} etc. There are several other theoretical
studies on the possible effects of large magnetic field in the initial
stage of heavy ion collisions. For example, flow harmonics and fluid
flow are shown to be altered in presence of non-zero magnetic field
along the perpendicular direction to the participant
plane~\cite{Pang:2016yuh,Inghirami:2016iru,Das:2017qfi,Greif:2017irh,Roy:2015kma,Pu:2016bxy}.
Transport coefficients of nuclear matter might also be changed under the influence 
of strong electro-magnetic fields \cite{McInnes:2017ohn,Tawfik:2016ihn,Hattori:2016cnt}.

It is now widely accepted view that due to the relativistic speed of
the charged protons inside the colliding nucleus a magnetic field as
large as $\sim 10^{19}$ Gauss is produced in mid-central Au+Au
collisions at $\sqrt{s_{\rm NN}}=$ 200 GeV
\cite{Deng:2012pc,Bzdak:2011yy,Roy:2015coa}. The magnitude of this initial
electromagnetic field increases almost linearly as a function of
$\sqrt{s_{\rm NN}}$ \cite{Deng:2012pc}.  However, the subsequent
space-time evolution of the initial magnetic field inside the Quark
Gluon Plasma (QGP - deconfined state of partons created in the
heavy-ion collisions~\cite{Gyulassy:2004zy,Adams:2005dq}) 
is still uncertain. The main uncertainty comes from the poorly known
value of the electrical conductivity $\sigma_{e}$ in the
QGP~\cite{Gupta:2003zh,Qin:2013aaa,Arnold:2000dr} phase 
and in the subsequent hadronic phase~\cite{Greif:2016skc,Steinert:2013fza,Ghosh:2016yvt}.

The space-time evolution of QGP (which contains electrically charged
quarks) in the presence of magnetic field is governed by the
relativistic magneto-hydrodynamic (MHD)
equations. Depending on the value of magnetic Reynolds number 
$R_m=\sigma_{e}vl\mu$, where $l$ is some characteristic length scale of the QGP;
$v$ is the fluid velocity; $\mu$ is the magnetic permeability of the QGP, 
one can approximate the MHD evolution as ideal (when $R_{m}\gg 1$) or 
resistive ($R_{m} \ll 1$) \cite{Goedbloed:2008}.  To get an order of
magnitude estimate of $R_{m}$, let us consider that in heavy ion collisions at top RHIC energy the typical fluid 
velocity is $v \sim 0.5 c$, the QGP possesses large electrical
conductivity $\sigma_{e} \sim 6T/e^2$ \cite{Arnold:2000dr} at
temperature $T=300$ $MeV$, assuming $\mu=1$ and  $l=10$ $fm$ we have
$R_{m}\sim 560$.  Hence as $R_{m}\gg 1$ one can treat the QGP evolution within the framework of ideal MHD. 
However, note that it is still an open question whether the QGP 
is in ideal MHD regime or in the resistive regime, for example 
\cite{Tuchin:2013apa} uses resistive MHD limit.

We know that in the initial stage of heavy ion collisions (particularly in non-central
collisions) the colliding nucleus carry large angular momentum~\cite{Becattini:2007sr}. A fraction of 
this large angular momentum might give rise to finite fluid vorticity~\cite{Becattini:2015ska,Pang:2016igs,Li:2017dan}.
The existence of strong magnetic fields and large flow
vorticity in the reaction zone may have interesting consequences.
In fact, in classical physics, the Larmor's theorem states that the motion of 
a charged particle in a magnetic field $\vec{B}$ is equivalent
to the motion in a rotating frame with angular velocity $\omega= q\vec{B}/(2m) $ with an additional
centrifugal force, where $m$ and $q$ are the mass and charge of the particle under consideration. 

Magnetic field and fluid vorticity can be coupled together and they follow a generalized form of advection equation 
\cite{Mahajan:2003,Mahajan:2010}. The origin of this generalized form of coupled magneto-vortical evolution 
trace back to the fact that the equations governing fluid vorticity and magnetic field
is of common form and we can take a linear combination of both of them \cite{Steinhauer:1997}. 
It is a well known fact that in the limit of vanishing electrical 
conductivity, plasma retains the initial magnetic flux passing through it, this phenomenon 
is known as frozen of magnetic flux. In this case the electric and
magnetic fields are no longer independent entities 
%once we fix the fluid three velocity $\vec{v}$, in fact they 
and are related as $\vec{E}=-\vec{v}\times\vec{B}$. 
The relativistic formulation of the problem requires a proper choice of 
fluid vorticity. 
%At this point we would like to stress that in the relativistic case 
The fluid vorticity could take many different 
forms, one of them is the thermal vorticity
$\omega^{\mu\nu}=\partial^{\nu}\left(fu^{\mu}\right)-\partial^{\mu}\left(fu^{\nu}\right)$
\cite{Luciano:2013,Lich:1967}, where $f(T)$ is in general a function of 
temperature, $u^{\mu}=(\gamma,\gamma\vec{v})$ is the fluid 4-velocity, and $\gamma=1/\sqrt{1-v^{2}}$ being
the relativistic kinematic factor (see Appendix \ref{app:2} for details). 
Note that $\omega^{\mu\nu}$ is actually different 
than its non-relativistic counterpart 
$\vec{\omega}= \vec{\nabla}\times \vec{v}$. From now on we shall refer the thermal 
vorticity as vorticity unless stated otherwise.

In this work we consider an ideal (zero viscosity and 
vanishing electrical resistivity) QGP expanding
under external homogeneous time varying magnetic field.
Since the electrical ressitivity of the fluid is very small we
use ideal MHD approximation. The aim is to study the temporal
evolution of the generalized magnetic field (initial magnetic field
coupled to vorticity) within a relativistic framework. We note that the authors of \cite{Deng:2012pc} 
have attempted to address similar problem for a 2+1 dimensional expansion of QGP under 
external magnetic field. However our work is different as it incorporates the both magnetic field and fluid vorticity effects. In
addition we have also considered relativistic correction to the evolution equation. 

It is worthwhile to point out that in order to investigate 
the space-time evolution of plasma and associated electromagnetic 
field one should solve the complete set of relativistic magnetohydrodynamics
equations with a proper initial condition and appropriate Equation 
of State (EoS). Recently few group have started investigating the 
magneto-hydrodynamical evolution of QGP by numerically solving the 
corresponding energy momentum tensor~\cite{Pang:2016yuh,Inghirami:2016iru,Das:2017qfi,Greif:2017irh,Roy:2015kma},
but none of the existing numerical codes 
solve the magnetohydrodynamics equation self consistently and at the 
same time consider non-zero fluid vorticity. 

The rest of the paper is organized as follows: in the next section 
we discuss the theoretical formulation and results. Section III
contains summary of our findings. We have re-derive some useful 
formulae associated with this study and those are given in the Appendix \ref{app:A} and \ref{app:2} for completeness.
Throughout the paper we use natural unit ($\hbar=c=k_{B}=1$ 
and the metric tensor $g^{\mu\nu}$= diagonal(1,-1,-1,-1). Four vectors 
are denoted by Greek indices, three vectors are denoted by arrow.

\section{\label{sec:level2}Formulation and result}
%%%%%%%%%%
%\iffalse
To set the stage one must begin with earliest non relativistic 
induction equation of MHD 
\begin{equation}\label{eqn:induMHD}
    \frac{\partial\widetilde{B}}{\partial t}-\vec{\nabla}\times\left(\vec{v}\times \widetilde{B}\right)=-\vec{\nabla}{n}^{-1}\times \vec{\nabla}p=\vec{\nabla}T\times\vec{\nabla}\left(\frac{s}{n}\right)
\end{equation}
where the term $\widetilde{B}=\vec{\omega}+q\vec{B}$ is called the generalised magnetic field (or generalised vorticity)
combines the fluid vorticity $\vec{\omega}=\vec{\nabla}\times\vec{v}$ and magnetic field $\vec{B}=\vec{\nabla}\times \vec{A}$
(where $\vec{A}$ is the usual vector potential) together in a single equation. 
Note that in the right hand side of Eq.(\ref{eqn:induMHD}) we have used the thermodynamic relation
\begin{equation}
 d\left(\frac{h}{n}\right)=Td\left(\frac{s}{n}\right)+\frac{1}{n}dp
\end{equation}
where $h$ is the enthalpy density, $s$ is the entropy density, $p$ is pressure and $T$ is the temperature
of the fluid and $n$ is the number density \cite{Landau:1987}. If the plasma is barotropic , i.e., a fluid whose pressure is a 
function of the number density only, i.e., $p = p(n)$, or in other words 
the entropy density is a function of only temperature, i.e., $s=s(T)$ the right side of
Eq.(\ref{eqn:induMHD}) is zero and the resulting ideal transport equation becomes  
\begin{equation}\label{eqn:NRfrozren}
 \frac{\partial\widetilde{B}}{\partial t}-\vec{\nabla}\times\left(\vec{v}\times \widetilde{B}\right)=0. 
\end{equation}
Both Kelvin's circulation theorem in hydrodynamics and the frozen-in flux theorem  of MHD, for example, 
are contained in Eq.(\ref{eqn:NRfrozren}) which states that for any loop $L(t)$ 
(or the enclosing surface $A(t)$) transported with the fluid, 
the circulation (or the flux) is conserved, i.e.,

    \begin{equation}
    \label{eq:NRKelvin}
\frac{d}{dt}\oint_{L(t)}\widetilde{P}\cdot d\vec x=\frac{d}{dt}\int_{A(t)}\widetilde{B}\cdot\hat{n}d^2x=0
\end{equation}
where $\widetilde{P}=\vec v+ q\vec A$. Note that the material derivative $d/dt=\partial /\partial t 
+\vec{v}\cdot\vec{\nabla}$.

But the same is not true for relativistic fluids. In principle we can combine the electromagnetic field tensor 
$F^{\mu\nu}$ and the flow field tensor ${\omega}^{\mu\nu}$, into an unified tensor 
$M^{\mu\nu}=qF^{\mu\nu}+\omega^{\mu\nu}$ (see Appendix {\ref{app:2}}). Generalizing Kelvin's theorem 
and the frozen-in flux theorem of MHD to relativistic fluids, forces us to switch to proper
time loop $L(\tau)$ (where proper time $\tau=t/\gamma$) in Eq.(\ref{eq:NRKelvin}). This transformation yields a Jacobian ${\gamma}^{-1}$ which makes the right hand side nonzero and we have 
(see Appendix {\ref{app:2}})
\begin{equation}
\frac{\partial \hat{B}}{\partial t}-\vec{\nabla} \times \left(\vec{v} \times \hat{B} \right)
=\frac{\gamma T}{2q}\vec{\nabla} v^2\times \vec{\nabla} \left(\frac{s}{n}\right)-\frac{1}{q\gamma}\vec{\nabla} T\times\vec{\nabla}\left(\frac{s}{n}\right),
\end{equation}
where the generalized magnetic field in the relativistic scenario is 
$\hat{B}=\vec{B}+\vec{\nabla}\times\left(f\gamma\vec v\right)$, $\vec{B}$ is the usual
magnetic field, $f=\frac{h}{n}$ is the enthalpy density ($h$) per unit conserved charge ($n$), $q$ is the absolute
magnitude of the electric charge of the fluid particles. Note that for $s(T)$ (which is generally true 
for hot and dense QGP) the second term on the right hand side yields zero but the first term might be
non-zero. Thus for ideal relativistic barotropic fluid we have
\begin{equation}
\label{eq:GenMagEvo}
  \frac{\partial \hat{B}}{\partial t}-\vec{\nabla} \times \left(\vec{v} \times \hat{B} \right)
= \frac{\gamma T}{2q}\vec{\nabla} v^2\times \vec{\nabla} \left(\frac{s}{n}\right) 
\end{equation}
The term in the right hand side of Eq.(\ref{eq:GenMagEvo}) has its origin in 
the space-time distortion caused by the demands of special relativity and is 
non existent in the non relativistic formulation. 
Following conclusion can be readily obtained from Eq.(\ref{eq:GenMagEvo}):
\begin{enumerate}
   \item The relativistic induced term on the right of Eq.(\ref{eq:GenMagEvo}) is zero for homogeneous
    entropy density  or for homogeneous flow fields.  
   \item It can also be zero if the term $\vec{\nabla}v^2$ is co-linear to $\vec{\nabla}s$.
\end{enumerate}

Now we shall discuss an application of the above mentioned generalized 
relativistic case to high energy heavy ion collisions.  
We consider a 2+1 dimensional ideal MHD 
evolution of QGP. The velocity  $v_{z}=\frac{z}{t}$ 
follow Bjorken one dimensional expansion
along the longitudinal direction (along the beam axis). 
Following \cite{Ollitrault:2008zz} we assume that the transverse expansion 
is slower compared to the longitudinal 
expansion. The fluid velocity in the transverse plane $\vec{v}_{\perp}=\hat{x}v_{x}+\hat{y}v_{y}$
is obtained from the following equation 
\begin{equation}
\label{eq:EoM}
    \frac{\partial \vec{v}_{\perp}}{\partial t}=-c_{s}^{2}\nabla_{\perp}s
\end{equation}
where, $c_{s}$ is the speed of sound, $s$ is entropy density of the fluid.
For the present work we assume a Gaussian transverse entropy(or energy density)
profile  
\begin{equation}
\label{eq:entrdens}
    s\left(x,y\right)=s_{0}exp\left(-\frac{(x-{\mu}_{x})^2}{2\sigma_{x}^2}-\frac{(y-{\mu}_{y})^2}{2\sigma_{y}^2}\right)
\end{equation}
where $\sigma_{x}$ and $\sigma_{y}$ are the Gaussian width along $x$ and $y$ direction 
respectively, $s_{0}$ is the entropy density at $x={\mu}_{x},y={\mu}_{y}$, for 
simplicity we take ${\mu}_{x}={\mu}_{y}=0$. By using Eq.(\ref{eq:entrdens}) into 
Eq.(\ref{eq:EoM}) a straightforward calculation gives 
\begin{eqnarray}
\label{eq:vx}
v_{x}&=&\frac{c_{s}^2 x}{2\sigma_{x}^{2}}t\\
\label{eq:vy}
v_{y}&=&\frac{c_{s}^2 y}{2\sigma_{y}^{2}}t
\end{eqnarray}
and according to the Bjorken assumption 
\begin{equation}
\label{eq:vz}
   v_{z}=\frac{z}{t}
\end{equation}

The transverse and longitudinal displacement of the fluid 
can be readily obtained by integrating $\vec{v}$ with respect to time
\begin{eqnarray}
\label{eq:r(t)}
x(t)&=&x_{0}\exp\left(\frac{c_{s}^{2}\left(t^{2}-t_{0}^{2}\right)}{2\sigma_{x}^{2}}\right) \\
\label{eq:y(t)}
y(t)&=&y_{0}\exp\left(\frac{c_{s}^{2}\left(t^{2}-t_{0}^{2}\right)}{2\sigma_{y}^{2}}\right)\\
\label{eq:z(t)}
z(t)&=&z_{0}\frac{t}{t_{0}}
\end{eqnarray}
where $x_0, y_0$ and $z_0$ are the initial transverse and longitudinal positions 
of the fluid at time $t=t_{0}$.

As mentioned earlier we investigate the ideal MHD evolution of QGP with non-zero
fluid vorticity. Note that according to the non-relativistic definition of vorticity 
$\vec{\omega}=\vec{\nabla}\times \vec{v}$, the fluid velocity 
$\vec{v}=\left(\frac{c_{s}^2 x}{2\sigma_{x}^{2}}t,
\frac{c_{s}^2 y}{2\sigma_{y}^{2}}t,\frac{z}{t} \right)$ 
gives $\vec{\omega}=\left(0,0,0\right)$. However, considering the relativistic 
generalisation of vorticity 
$\omega^{\mu\nu}=\partial^{\nu}\left(fu^{\mu}\right)-\partial^{\mu}\left(fu^{\nu}\right)$
(see Appendix \ref{app:2}) gives non-zero value for the given velocity profile.
In addition to that the frozen-in flux theorem gets modified 
in the relativistic regime.

%%%%%%%%%%%%%%%%%%%%%%%%%%%%%
We take boost invariance along longitudinal direction and hence 
fluid properties depends only in the transverse direction near the 
mid rapidity. Note that the term on the right hand side
of Eq.(\ref{eq:GenMagEvo}) appears due to the relativistic effects and
non-zero vorticity, thus in the non-relativistic
limit and vanishing vorticity  one gets back the usual form of ideal MHD evolution of 
magnetic field. 
\begin{equation}
\label{eq:FrozenFluxUsual}
\frac{\partial \vec{B}}{\partial t}=\vec{\nabla} \times \left(\vec{v} \times  \vec{B} \right)
\end{equation}
Using the above equation \cite{Deng:2012pc} have found analytic solution 
of magnetic field evolution for 2+1 dimensional expansion of QGP in the ideal 
MHD limit which is discussed in Appendix \ref{app:A}.
Let us now consider Eq.(\ref{eq:GenMagEvo}) and investigate the 
consequence of non-zero vorticity, finite gradient of entropy
density and fluid velocity in the relativistic generalization. 

\begin{figure*}
 \center
  \includegraphics[width=0.6\textwidth]{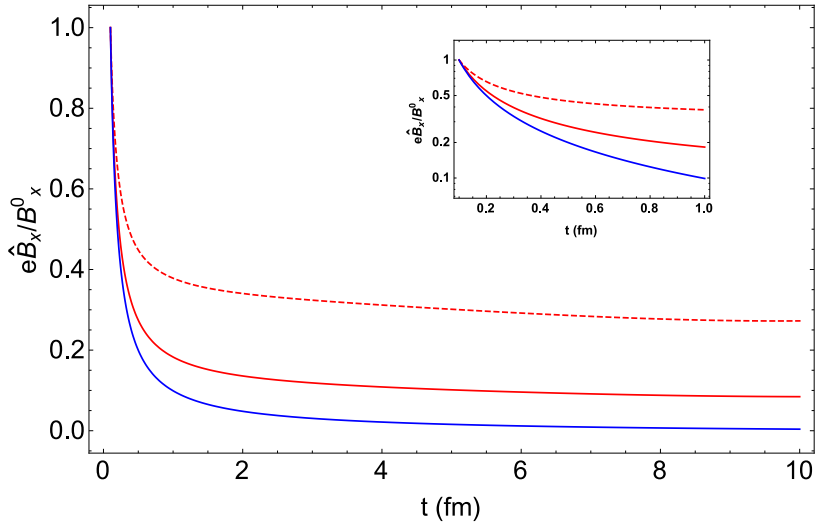}
  \caption{Temporal evolution of normalized $\hat{B}_{x}$ for z$>$0. Blue line
  corresonds to the usual magnetic field evolution, red solid line
  corresponds to $\hat{B}_x$ for $\gamma=3$, whereas red dashed line
  corresponds to $\hat{B}_x$ for $\gamma=10$. The inset figure shows the same
  result but with log scale.}
  \label{fig:ZpB_xGamma3}
\end{figure*}

\begin{figure*}
 \center
  \includegraphics[width=0.6\textwidth]{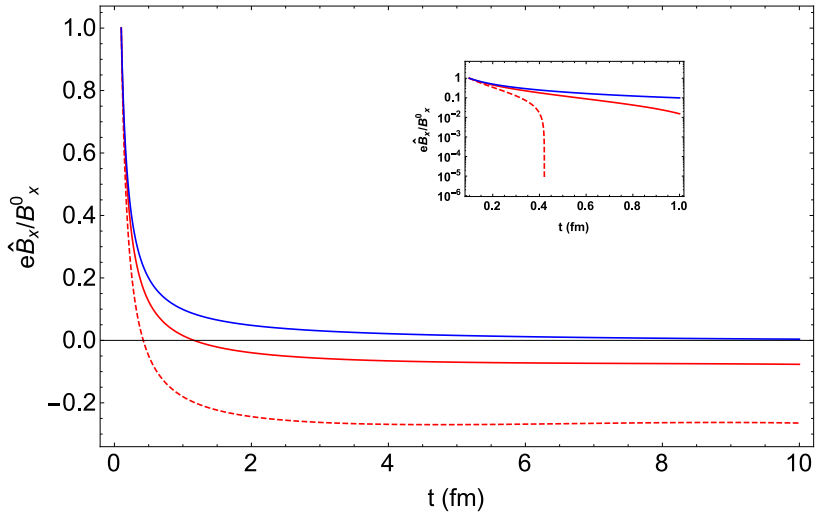}
  \caption{Same as Fig.\ref{fig:ZpB_xGamma3} but for z$<$0.}
  \label{fig:ZnB_xGamma3}

\end{figure*}

%%%%%%%%%%%%%%%%%%%%%%%%%%%%%%%%%%
% Ashutosh starts here 

For simplicity let us consider that the net conserved charge $n$ is 
homogeneous and constant over the transverse plane. 
Note that $\vec{\nabla} v^2\times \vec{\nabla} s$= $s\left(\frac{2zy}{t^2\sigma_y^2} 
,-\frac{2zx}{t^2\sigma_x^2},-2\left(\frac{xyg_{x}t^{2}}{\sigma_{y}^{2}}-
\frac{xyg_{y}t^{2}}{\sigma_{x}^{2}}\right)\right)$,
where $g_{x}=c_{s}^{4}/\sigma_{x}^{4}$ and $g_y=c_{s}^{4}/\sigma_{y}^4$.
Thus from Eq.(\ref{eq:GenMagEvo}) we have the evolution equation of the component of $\hat{B}$ 
\begin{eqnarray}
\frac{\partial \hat{B_x}}{\partial t}+\left(\frac{c_s^2t}{\sigma _y^2}+\frac{1}{t}\right)\hat{B_x} &=& \left(\frac{\gamma T}{2q}\right)\frac{2zy}{nt^2\sigma_y^2}s
\label{BhatX} \\
\frac{\partial \hat{B_y}}{\partial t}+\left(\frac{c_s^2t}{\sigma _x^2}+\frac{1}{t}\right)\hat{B_y} &=& -\left(\frac{\gamma T}{2q}\right)\frac{2zx}{nt^2\sigma_x^2}s
\label{BhatY}
\end{eqnarray}

Notice the different sign on the right hand side of Eq.(\ref{BhatX}) and (\ref{BhatY}), 
which will give rise to completely opposite evolution for $\hat{B}_x$
and $\hat{B}_y$. The solution also depends on the forward or backward 
rapidity ($z>0$,$z<0$).
The solution of Eq.(\ref{BhatX}) and (\ref{BhatY}) can be found analytically.
First note that setting right hand side of Eq.(\ref{BhatX}) and (\ref{BhatY}) to zero
we have the homogeneous equations and the corresponding 
integrating factors are $M(t)=exp\left(\frac{c_s^2t^2}{2\sigma _x^2}\right)t$, and
$N(t)=exp\left(\frac{c_s^2t^2}{2\sigma _y^2}\right)t$ respectively.
Multiplying Eq.(\ref{BhatX}) and Eq.(\ref{BhatY}) by the corresponding integrating factors 
we have 
\begin{eqnarray}
\label{eq:BhatX2}
\frac{\partial }{\partial t}\left(\hat{B_{x}}N(t)\right) &=& \left(\frac{\gamma T}{2q}\right)\frac{2szyN(t)}{nt^2\sigma_y^2} \\
\label{eq:BhatY2}
\frac{\partial }{\partial t}\left(\hat{B_{y}}M(t)\right)&=&-\left(\frac{\gamma T}{2q}\right)\frac{2szxM(t)}{nt^2\sigma_x^2} 
\end{eqnarray}
At this point we should use the explicit time dependence of $x(t), y(t), z(t), s(t)$, and $n(t)$ in order to solve 
Eq.(\ref{eq:BhatX2}) and (\ref{eq:BhatY2}). 
$x(t),y(t)$ and $z(t)$ are given in Eq.(\ref{eq:r(t)}-\ref{eq:z(t)}).  

Here we shall use another assumption in order to evaluate $s(t)$ and $n(t)$.
Initially the longitudinal expansion is much faster compared to the transverse expansion
the entropy density evolves as 

%%%%%%% image

\begin{equation}
\label{eq:s(t)}
 s(t)=\frac{s_0 t_0}{t}
\end{equation}
Similarly the number density $n(t)$ is given as 
\begin{equation}
 n(t)=\frac{n_0t_0}{t}.
\end{equation}
In reality the time dependence may be somewhat different than given in Eq.(\ref{eq:s(t)}), we envisage 
that will not change the qualitative nature of the solution we obtained here.
Inserting the time dependence of the above mentioned quantities in Eq.(\ref{eq:BhatX2}) and (\ref{eq:BhatY2})
yields
\begin{eqnarray}
\label{eq:finalBhat1}
 \frac{\partial }{\partial t}\left(\hat{B}_{x} N(t)\right)&=&\left(\frac{\gamma T}{q}\right)C_1\exp\left(\frac{c_s^2(2t^2-t_0^2)}{2\sigma_y^2}\right) \\
 \label{eq:finalBhat2}
 \frac{\partial }{\partial t}\left(\hat{B}_{y} M(t)\right)&=&-\left(\frac{\gamma T}{q}\right)C_2\exp\left(\frac{c_s^2(2t^2-t_0^2)}{2\sigma_x^2}\right)
\end{eqnarray}
where $C_1=\frac{s_0z_0 y_0}{n_0t_0\sigma_y^2}$, $C_2=\frac{s_0z_0 x_0}{n_0t_0\sigma_x^2}$.

Integrating Eq.(\ref{eq:finalBhat1}) and (\ref{eq:finalBhat2}) with respect to time yields  
\begin{widetext}
 \begin{eqnarray}
 \label{eq:GenBhat1}
 \hat{B}_{x}=C_{1}\left(\frac{\gamma T}{q}\right)\frac{\sqrt{\pi}\sigma_y}{c_s t}\left[\erfi\left(\frac{c_s t}{\sigma_y}\right)-\erfi\left(\frac{c_s t_0}{\sigma_y}\right)\right]
 e^{-\frac{c_s^2(t^2-t_0^2)}{2\sigma _y^2}} + \frac{B^0_xt_0}{t}e^{-\frac{c_s^2(t^2-t_0^2)}{2\sigma _y^2}} \\
 \label{eq:GenBhat2}
 \hat{B}_{y}=-C_{2}\left(\frac{\gamma T}{q}\right)\frac{\sqrt{\pi}\sigma_x}{c_s t}\left[\erfi\left(\frac{c_s t}{\sigma_x}\right)-\erfi\left(\frac{c_s t_0}{\sigma_x}\right)\right]
  e^{-\frac{c_s^2(t^2-t_0^2)}{2\sigma _x^2}} + \frac{B^0_yt_0}{t}e^{-\frac{c_s^2(t^2-t_0^2)}{2\sigma _x^2}}
 \end{eqnarray}
\end{widetext}

Eqs. (\ref{eq:GenBhat1}) and (\ref{eq:GenBhat2}) are the main result of this study. They describe the 
temporal evolution of $\hat{B}$ which includes both usual magnetic field (last term on the right hand side of 
both equations) and the relativistic 
correction (first term on the right hand side of both equations) 
arising due to non-zero vorticity and entropy gradient. Note that for a given $z$
the relativistic correction has opposite sign for $\hat{B}_{x}$ and $\hat{B}_y$.
This is reflected in the numerical result shown in Fig.(\ref{fig:ZpB_xGamma3})
and (\ref{fig:ZnB_xGamma3}).
Where we use the following values 
$s=100$ $MeV^3$,
$n=0.4$ $MeV^3$,
$T=200$ $MeV$, 
$\gamma=10$ (dashed lines) and $\gamma=3$ (solid lines),
initial transverse coordinates $x_0=y_0=0.1$ $fm$ , $t_0=0.1$ $fm$,
$\sigma_x=\sigma_y=4.2$ $fm$, $c_{s}^{2}=1/3$. We used these
particular values of the parameter set because 
the experimental data in Au+Au collisions at $\sqrt{s_{NN}}=200$ GeV 
in 2+1D hydrodynamics simulation are best described by similar values
of the parameter set \cite{Kolb:2003dz}.

Since the final result
depends on the sign of $z$ we choose $z_0=0.1$ $fm$ and $z_0=-0.1$ $fm$ for 
showing the dependence. The initial value of the magnetic field 
$B^0_x$, $B^0_y$ was set to $m_{\pi}^2$. In Fig.(\ref{fig:ZpB_xGamma3})
and Fig.(\ref{fig:ZnB_xGamma3}) the blue lines correspond to usual magnetic 
field $B_{x}$ and red line corresponds to generalized magnetic field $\hat{B}_x$.
Fig.(\ref{fig:ZpB_xGamma3}) is for $z_0>0$ and Fig.(\ref{fig:ZnB_xGamma3})
corresponds to  $z_0<0$. If we assume that the effective magnetic field 
$\hat{B}$ acts as usual magnetic field then the result implies that the 
time evolution of magnetic field depends on the rapidity as well as 
the component of generalised magnetic field along or perpendicular to the reaction plane 
(spanned by $x$ and $z$ axis). This then gives rise to rapidity dependence of the predicted 
chiral magnetic effect and other related phenomenon. 

\section{Summary}
In this work we explored the space-time evolution of finite fluid vorticity 
in presence of non-zero magnetic fields in a 2+1 dimensional ideal
fluid expansion with Bjorken boost invariance in heavy ion collisions. 
Throughout the work we use the fact that even in the ideal magnetohydrodynamics
regime finite fluid vorticity might arises in presence of  inhomogeneous 
flow fields and the fluid entropy density. Using an ideal EoS for massless quarks and gluons
we found analytic solution (Eqs. (\ref{eq:GenBhat1}) and (\ref{eq:GenBhat2})) of generalised 
magnetic field. The solutions have 
some non-trivial dependence on the position of the fluid with respect to midplane (z=0).
If one assume that $\hat{B}$ acts similarly as $\vec{B}$ then our finding suggest 
the temporal evolution of magnetic field in heavy ion collisions 
will be changed in presence of finite fluid vorticity. 
We would like to point out that several assumptions have been made in this study;
like we assume an idealistic fluid velocity field, zero shear and bulk viscosity,
an ideal gas EoS, and we have neglected the back reaction of the magnetic field to the 
fluid velocity, in the next step (possibly in a future work) 
we need to improve upon these assumptions.

\acknowledgements
We would  like to thank S Mahajan for helpful discussion.
V.R. is supported by the DST-INSPIRE faculty research grant.

%%%%%%%%%%%%%%%%%%%%%%%%%%%%%%%%%%%
%%%%%% Appendix starts here 
%%%%%%%%%%%%%%%%%%%%%%%%%%%%%%%%%%%
\appendix

\section{Ideal MHD evolution of QGP in 2+1 dimension.}
\label{app:A}
Here we show how to obtain the result found in \cite{Deng:2012pc} by Deng et al 
from the general formula Eq.(\ref{eq:GenMagEvo}).
This can be done by setting the right hand side of Eq.(\ref{eq:GenMagEvo}) to zero and noting that the
generalised magnetic field $\hat{B}$ is replaced by the usual magnetic field $\vec{B}$. Thus, we have
\begin{equation}
\label{eq:App1B}
\frac{\partial  \vec{B}}{\partial t}-\vec{\nabla} \times \left(\vec{v} \times  \vec{B} \right)=0.
\end{equation}

The components of $\vec{v}$ are given in Eq.(\ref{eq:vx} - \ref{eq:vz}).
We have from Eq.(\ref{eq:App1B}) 
\begin{equation}\label{5}
\frac{\partial B_y}{\partial t}=-\left(\frac{c_s^2t}{\sigma _x^2}+\frac{1}{t}\right)B_y.
\end{equation}
Solving the above equation with the initial value of magnetic field at time $t=t_0$ to be $B_y^0$ gives us
\begin{equation}\label{6}
B_y=\frac{t_0}{t}exp\left(-\frac{c_s^2(t^2-t_0^2)}{2\sigma _x^2}\right)B^0_y.
\end{equation}
This is the desired result as was obtained in \cite{Deng:2012pc}.

%%%%%%%%%%%%%%%%%%%%%%%%%%%%%%%%%%%%%%%
%%%   APPENDIX -2 
%%%%%%%%%%%%%%%%%%%%%%%%%%%%%%%%%%%%%%%

\section{Formulation of relativistic magneto-vortical evolution}
\label{app:2}
Here we shall discuss important formulation used in our present work, most of them can be found 
in the original references \cite{Mahajan:2010,Mahajan:2003}.
We begin our derivation by writing few general equations for co-variant relativistic magnetohydrodynamics (MHD). 
We use the general definition of energy momentum tensor of fluid and the particle four current 
\begin{eqnarray}
T^{\mu\nu}&=&hu^{\mu}u^{\nu}-pg^{\mu\nu}
\label{A.8}
\\
\Gamma^{\mu}&=&nu^{\mu}
\label{A.9}
\end{eqnarray}
where $h$ is the enthalpy density, $p$ is the fluid pressure and $n$ is the baryon density 
(or conserved charge density) in the rest frame of the given fluid. 
The evolution of electromagnetic field is governed by Maxwell's equation and the 
four current conservation equations 
\begin{eqnarray}
\partial_{\nu}F^{\mu\nu}&=&J^{\mu}, 
\label{eq:AppBFieldTensor}
\\
\partial_{\nu}\Gamma^{\nu}&=&0.
\label{eq:AppBJnu}
\end{eqnarray}
The current $J^{\mu}=q\Gamma^{\mu}$ couples the fluid to the Maxwell's equations 
giving us the equation of motion as
\begin{equation}\label{eq:A.12}
\partial_{\nu}T^{\mu\nu}=F^{\mu\nu}J_{\nu}.
\end{equation}
We also define $f(T)$ which is a function of temperature only, 
$f(T)$ is assumed to be factorized in terms of the conserved charge density ($n$)
and enthalpy density ($h$) as 
\begin{equation}
\label{eq:appBfactF}
f(T)=\frac{h}{n}.
\end{equation}
Now let us introduce a completely anti-symmetric second rank fluid tensor 
$\omega^{\mu\nu}$ sometimes also called the \textit {thermal vorticity tensor} \cite{Becattini:2015ska}.
\begin{equation}\label{A.14}
\omega^{\mu\nu}=\partial^{\nu}\left(fu^{\mu}\right)-\partial^{\mu}\left(fu^{\nu}\right).
\end{equation}
More specifically one can write $\omega^{\mu\nu}=(-\vec Q,\vec R)$, thus we see that $\omega^{\mu\nu}$
is like the electromagnetic field tensor $F^{\mu\nu}$ and the vectors $(-\vec Q,\vec R)$ correspond
to $(\vec{E},\vec{B})$. The explicit expression of $\vec{Q}$ and $\vec{R}$ are 

\begin{eqnarray}
\vec{Q}&=&\vec{\nabla}\left(f\gamma\right)\nonumber-\frac{\partial}{\partial t}\left(f\gamma\vec{v}\right)\\
\vec{R}&=&\vec{\nabla}\times\left(f\gamma\vec v\right).
\end{eqnarray}

We note that the fluid tensor $\omega^{\mu\nu}$ is exactly equivalent to
the Maxwell's tensor $F^{\mu\nu}$, for example like the non-existence of 
magnetic monopoles in the Maxwell's theory ($\vec{\nabla}\cdot\vec{B}=0$) 
we have $\vec{\nabla}\cdot\vec{R}=0$ in the fluid case.

Using the function $f(T)$ Eq.(\ref{eq:appBfactF}) in Eq.($\ref{eq:A.12}$) one gets
\begin{equation}\label{A.15}
nu^{\mu}\partial_{\mu}\left(fu^{\nu}\right)-\partial^{\nu}p=qnF^{\nu\mu}u_{\mu}
\end{equation}
where we have used Eq.($\ref{eq:AppBJnu}$) to reach ($\ref{A.15}$).\par
We define a new unified field tensor $M^{\mu\nu}$ from the combination of 
$F^{\mu\nu}$ and $\omega^{\mu\nu}$ as
\begin{align}\label{A.18.5}
M^{\mu\nu} &=qF^{\mu\nu}+\omega^{\mu\nu}\nonumber\\
&=qF^{\mu\nu}+\partial^{\nu}\left(fu^{\mu}\right)-\partial^{\mu}\left(fu^{\nu}\right).
\end{align}

We contract Eq.($\ref{A.18.5}$) with $u^{\mu}$ to get
\begin{equation}\label{A.16}
u^{\mu}\partial_{\mu}\left(fu^{\nu}\right)=\partial^{\nu}f+u_{\mu}\left(qF^{\mu\nu}-M^{\mu\nu}\right)
\end{equation}
Substituting Eq.(\ref{A.16}) into Eq.(\ref{A.15}), we have
\begin{align}
&n\partial^{\nu}f-\partial^{\nu}p=nu_{\mu}M^{\mu\nu}\label{A.17}
\\
&T\partial^{\nu}\left(\frac{s}{n}\right)=u_{\mu}M^{\mu\nu}\label{A.18}
\end{align}
where we have used the basic thermodynamic relation $df=Td(s/n)+(1/n)dp$ in Eq.($\ref{A.17}$).\par

The canonical momentum of a charge $q$ of unit mass in presence of a magnetic field 
($\vec{\nabla}\times\vec{A}$) with velocity $\vec v$ 
is $\vec P=\vec v+ q\vec A$. The corresponding generalized vorticity is 
$\vec{\nabla}\times \vec P$. 
In the non-relativistic limit the fluid canonical momentum satisfies 
the well known circulation theorems 
\begin{equation}\label{A.19}
\frac{d}{dt}\oint_{L(t)}\vec P\cdot d\vec x=\oint_{L(t)}\left(\frac{\partial\vec P}{\partial t}+\vec{\nabla}\times\left(\vec P\times \vec{v}\right)\right)\cdot d\vec x,
\end{equation}
and is zero if the right hand side is an exact differential. From the equation of motion for non relativistic fluids one sees that the right hand side of Eq.($\ref{A.19}$) is the gradient of effective energy $E$.

\begin{equation}\label{A.19.5}
 \frac{\partial\vec P}{\partial t}+\vec{\nabla}\times\left(\vec P\times \vec{v}\right)=-\vec{\nabla} E.
\end{equation}
Substituting Eq.($\ref{A.19.5}$) and using Stokes's theorem in Eq.($\ref{A.19}$) one immediately deduces 
the Kelvin's circulation theorem
\begin{equation}
 \frac{d}{dt}\int_{S(t)} \left({\vec{\nabla}\times\vec{P}}\right)\cdot d\vec S=\oint_{L(t)}\vec{\nabla} E\cdot dx=0.
\end{equation}
\par
The previous argument doesn't hold for relativistic fluids since in a co-variant picture one should talk about the proper time $\tau$ instead of the coordinate time $t$. Similarly the coordinate loop $L(t)$ must be replaced by relativistic loop $L(\tau)$. Updating to the above redefinition leads us to a relativistic circulation theorem along the loop $L(\tau)$ obeying \cite{Mahajan:2010,Mahajan:2003}
\begin{equation}\label{A.20}
\frac{d}{d\tau}\left(\oint_{L(\tau)}\Pu^{\mu}dx_{\mu}\right)=\oint_{L(\tau)}\left(\partial^{\mu}\Pu^{\nu}-\partial^{\nu}\Pu^{\mu}\right)u_{\mu}dx_{\nu},
\end{equation}
where $\Pu^{\mu}$ is the appropriate momentum relating the right hand side of Eq.($\ref{A.20}$) with an effective force, 
an exact differential. One can easily guess the relativistic generalized 4-momentum coupled to magnetic field through the four 
potential $A^{\mu}$ to be $\Pu^{\mu}=fu^{\mu}+qA^{\mu}$, and the integrand becomes
\begin{equation}\label{A.21}
 M^{\mu\nu}=\partial^{\mu}\Pu^{\nu}-\partial^{\nu}\Pu^{\mu}=qF^{\mu\nu}+\omega^{\mu\nu}   
\end{equation}
 the same unified tensor $M^{\mu\nu}$, that we have defined in Eq.($\ref{A.18.5}$). 
 Substituting $M^{\mu\nu}$ in Eq.($\ref{A.20}$) and using the equation of motion Eq.($\ref{A.18}$), 
 one finds an exact differential consisting the entropic driven force $T\partial^{\nu}\left(\frac{s}{n}\right)$ 
 \begin{equation}\label{A.22}
\frac{d}{d\tau}\left(\oint_{L(\tau)}\Pu^{\mu}dx_{\mu}\right)=\oint_{L(\tau)}T\partial^{\nu}\left(\frac{s}{n}\right)dx_{\nu}.
 \end{equation}
 Thus one arrives at the relativistic generalisation of Kelvin's circulation theorem. 
 However, note that the vorticity or the magnetic field is defined inside the coordinate loop $L(t)$, 
 thus we have to map back from relativistic loop $L(\tau)$ to coordinate loop $L(t)$. 
 But mapping back imparts a distortion owing to the Jacobian ${\gamma}^{-1}$ which spoils the exactness 
 (exact differential) of the right hand side which is the thermodynamic force.
 Taking the vector part of equation of motion Eq.($\ref{A.18}$) for relativistic fluids 
 \begin{equation}
 \label{A.23}
  q\left(\hat E+ \vec{v}\times\hat B \right)=\frac{T}{\gamma}\vec{\nabla}\left(\frac{s}{n}\right),
 \end{equation}
 the left hand side is a generalized statement of Lorentz's force, which is here equals to the entropy driven force defined 
 elsewhere in the text. \par
 \begin{align}
 &\hat{E}=\vec{E}+\vec{\nabla}\left(f\gamma\right)\nonumber-\frac{\partial}{\partial t}\left(f\gamma\vec{v}\right)\\
&\hat{B}=\vec{B}+\vec{\nabla}\times\left(f\gamma\vec v\right)
 \end{align}
 and it satisfies Faraday's law i.e., $\partial_t \hat B=-\vec{\nabla}\times \hat E$. Taking the curl of 
 Eq.($\ref{A.23}$) and using Faraday's law for $\hat{E}$ and $\hat{B}$ yields 
  \begin{widetext}
 \begin{eqnarray}
 \label{A.24}
  q\left(\frac{\partial\hat B}{\partial t}-\vec{\nabla}\times \left(\vec{v}\times\hat B\right) \right)&=&-\vec{\nabla}\times\left(\frac{T}{\gamma}\vec{\nabla}\left(\frac{s}{n}\right)\right)
    \\\nonumber
   &=&-\frac{1}{\gamma}\vec{\nabla} T\times\vec{\nabla}\left(\frac{s}{n}\right)+\frac{\gamma T}{2}\vec{\nabla} v^2\times \vec{\nabla} \left(\frac{s}{n}\right).
 \end{eqnarray}
  \end{widetext}
 This is the main formula used in our present work.
 For ideal barotropic fluid, i.e., a fluid whose pressure $P$ is a function of the number density 
 $n$, i.e., $P = P (n)$ the second term on the right hand side of 
 Eq.($\ref{A.24}$) will be dominant.
 This is because the thermodynamic coupling between entropy and temperature will make the 
 gradients of $s$ and $T$ co-linear, but notice that there is no such constraint which will 
 guarantee the colinearity of $\vec{\nabla}v$ and $\vec{\nabla}s$. Hence for our purpose we 
 will be only interested in the second term. Thus, we have
 \begin{equation}
 \label{eq:appE}
  \left(\frac{\partial\hat B}{\partial t}-\vec{\nabla}\times \left( v\times\hat B \right)\right)=
   \frac{\gamma T}{2q}\vec{\nabla} v^2\times \vec{\nabla} \left(\frac{s}{n}\right)
 \end{equation}
  Note that in the non-relativistic limit the second term on the right hand side of Eq.(\ref{A.24})
  does not exists, we only have the first term which is identical to zero for barotropic fluids (for example 
  QGP is also considered to be a barotropic fluid).

 %%%%%%%%%%%%%%%%%%%%%%%%%%%%%%%%%%%%
 % Bibliography
 %%%%%%%%%%%%%%%%%%%%%%%%%%%%%%%%%%%%

\end{document}